\documentclass[usenatbib,usegraphicx]{mn2e}
\usepackage{times}
\usepackage{psfig,latexsym,graphicx,natbib} 
\bibpunct{(}{)}{;}{a}{}{,}
\pssilent
\begin{document}
   \title[Magnetic field detection in the B2\,Vn star HR 7355]{Magnetic field detection in the B2\,Vn star HR 7355\thanks{Based on
       observations made with ESO Telescopes at the Paranal Observatory under
       programme ID: 081.D-2005}}

\author[Th. Rivinius et al.]
{Th.~Rivinius$^1$\thanks{triviniu@eso.org},
Th.~Szeifert$^1$,
L.~Barrera$^2$,
R.H.D.~Townsend$^3$,
S.~\v{S}tefl$^1$,
D.~Baade$^4$
\\
%
$^1$ESO - European Organisation for Astronomical Research in the Southern
Hemisphere, Casilla 19001, Santiago 19, Chile \\
$^2$UMCE \\
$^3$Department of Astronomy, University of Wisconsin--Madison, Sterling Hall,
475 N Charter St., Madison 53711, USA \\
$^4$ESO - European Organisation for Astronomical Research in the Southern Hemisphere, Karl-Schwarzschild-Str.~2, 85748 Garching bei M\"unchen, Germany 
}
\date{Accepted: $<$date$>$; Received: $<$date$>$; \LaTeX ed: \today}      
   \maketitle

\begin{abstract}
    {The B2Vn star HR\,7355 is found to be a He-rich magnetic
      star. Spectropolarimetric data were obtained with FORS1 at UT2 on
      Paranal observatory to measure the disk-averaged longitudinal magnetic
      field at various phases of the presumed 0.52\,d cycle. A variable
      magnetic field with strengths between $\langle B_z \rangle=-2200$ and
      $+3200$\,G was found, with confidence limits of 100 to 130\,G.  The
      field topology is that of an oblique dipole, while the star itself is
      seen about equator-on.  In the intensity spectra the {He}{\sc i}-lines
      show the typical equivalent width variability of He-strong stars,
      usually attributed to surface abundance spots. The amplitudes of the
      equivalent width variability of the {He}{\sc i} lines are
      extraordinarily strong compared to other cases.  These results not only
      put HR\,7355 unambiguously among the early-type magnetic stars, but
      confirm its outstanding nature: With $v\sin i = 320$\,km\,s$^{-1}$ the
      parameter space in which He-strong stars are known to exist has doubled
      in terms of rotational velocity. }
\end{abstract}
\begin{keywords}
Stars: individual: HR7355 --- Stars: hot, magnetic, chemically peculiar
\end{keywords}
\section{Introduction}
\citet{2008A&A...482..255R} suggested the B2Vn star {HR\,7355} (HD\,182\,180,
HIP\,95\,408) to be a He-rich magnetic star with a magnetosphere containing
trapped gas that produces hydrogen line emission
(\citealt{2005MNRAS.357..251T}, \citealt*{2007MNRAS.382..139T}). They based
their suggestion on evidence derived from two FEROS echelle spectra and a
Hipparcos light-curve, with a period of either single-wave 0.26\,d or
double-wave 0.52\,d. However, a period of 0.26\, could not be due to
rotational modulation, since this would require a rotational speed well above
the critical threshold.

If confirmed, HR\,7355 would have a unique position in the He-strong class:
With $v\sin i = 320$\,km\,s$^{-1}$ \citep*{2002ApJ...573..359A} it would be
the the most rapidly rotating magnetic star in the upper HR-diagram; by a
factor of about two ahead of the current record-holder, {$\sigma$\,Ori~E},
which has $v\sin i = 165$\,km\,s$^{-1}$ (\citeauthor{2002ApJ...573..359A},
op.~cit.).  This would make the star a show-case for the Rigidly Rotating
Magnetosphere model, that so far, though with great success, was applied to
only one star, $\sigma$\,Ori\,E (\citeauthor{2005MNRAS.357..251T}, op.~cit.).
In order to test this claim, a spectropolarimetric campaign was carried out in
2008.

\begin{figure}
\centering
\includegraphics[angle=270,width=0.95\columnwidth,clip=]{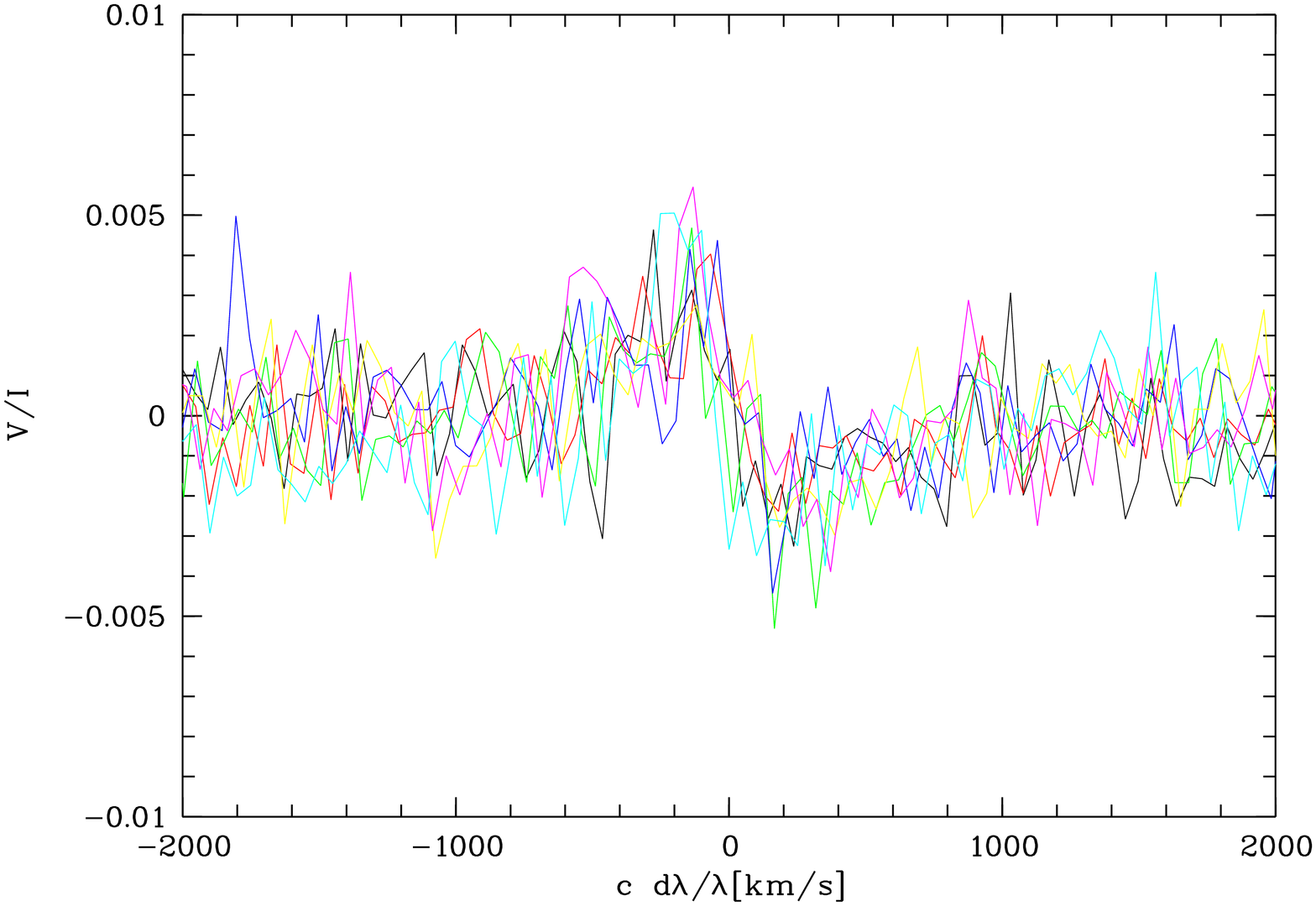}

\includegraphics[angle=270,width=0.95\columnwidth]{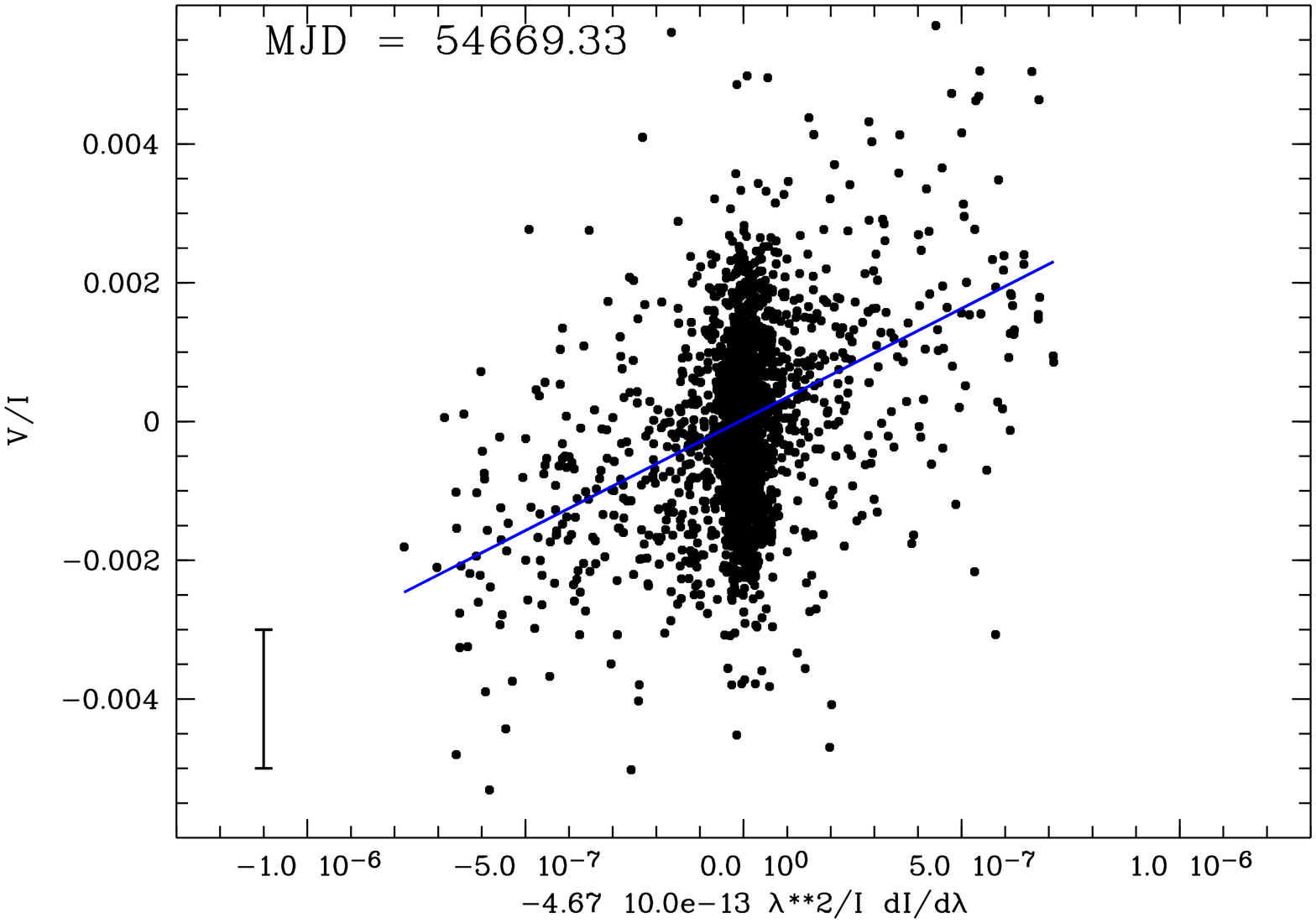}%

\caption[]{Normalized circular polarization $V/I$ of HR\,7355 at MJD=54669.33,
  for which a longitudinal magnetic field component of 3200\,G was derived. Top:
  The spectra around the ten Balmer lines from H$\beta$ to H$_{13}$. Bottom:
  The $V/I$ vs.\ the bracketed part of the right term in Eqn.~\ref{eqn_reg}
  and the linear regression to the data as a solid line. The typical error per
  measurement is shown as a bar; the error in the abscissa is less than the
  symbol size. The mass of points clustering around (0,0) are due to the
  unpolarized continuum.}
\label{fig_FORSV}
\end{figure}

\begin{figure}
\centering

\includegraphics[width=0.95\columnwidth,clip]{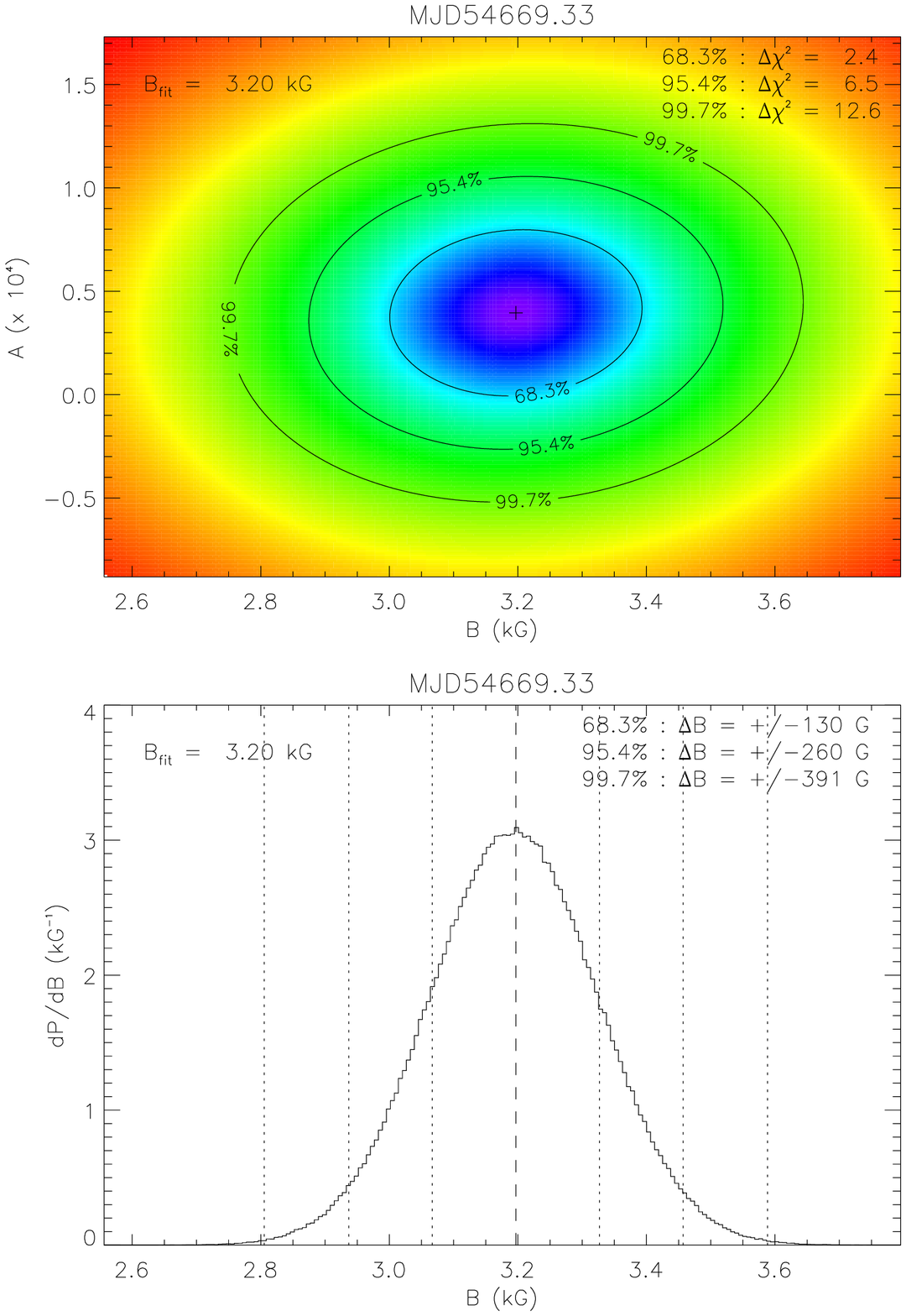}%

\includegraphics[width=0.95\columnwidth,clip]{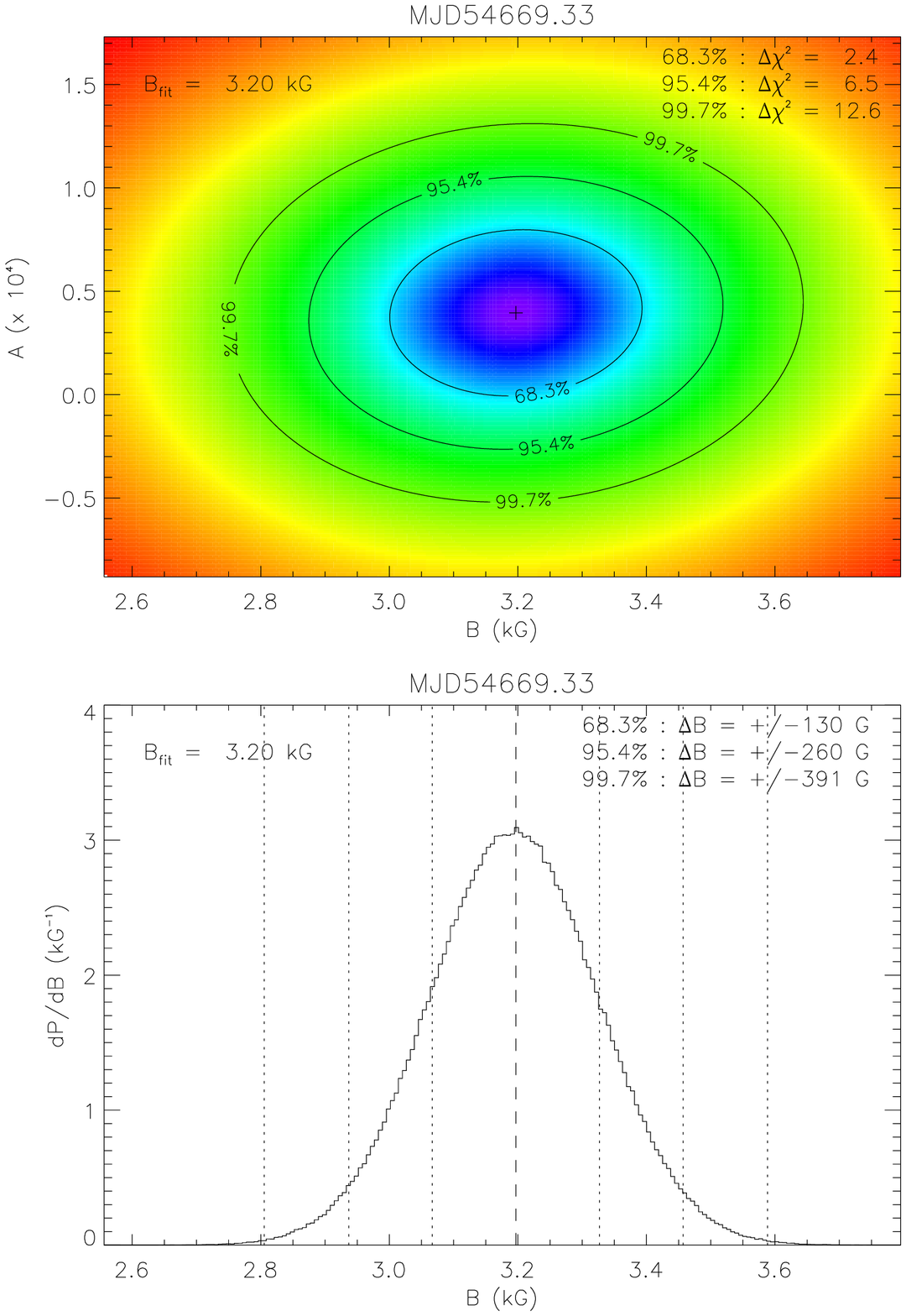}%

\caption[]{Minimum of the two-dimensional probability density disctribution
  computed from the artificial test data (see Sect.~\ref{sec_mcbs}) for the
  observed data shown in Fig.~\ref{fig_FORSV}, and an overplot of the
  $\chi^2$-contrours enclosing 68.3\%, 95.4\% and 99.7\% probabilities (upper
  panel) and one dimensional probability distribution of the derived field
  strength (lower panel).  }
\label{fig_FORSB}
\end{figure}

\section{Observations and Data Reduction}

The data described below were obtained at the VLT with the FORS1 instrument
\citep{1998Msngr..94....1A}, which is equipped with a super-achromatic
quarter-wave retarder plate. The instrument setup used the 1200B holographic
prism with a dispersion of 24\,\AA/mm over a range of 366 to 511\,nm. At a
slit-width of 0.3\arcsec, a (binned) pixel scale of 0.25\arcsec per pixel, and
a (binned) pixel size of 30\,$\mu$m, this gives about 0.9\,\AA\ per pixel,
which slightly oversamples the resolution element of about 1.3\,\AA.

Each individual measurement (see Table~\ref{tab_result}) consisted of a
sequence of eight exposures, taken at retarder plate position angles in the
sequence
($-45\degr,-45\degr,45\degr,45\degr,-45\degr,-45\degr,45\degr,45\degr$) with
respect to the axis of the Wollaston prism. The exposure time was two seconds
for the respective 8 exposures. The data was taken during four nights in
service mode. During two of these nights the sequence was repeated, so that we
have obtained six data sets, each one reaching an approximate S$/N$-ratio of
about 1000 per resolution element.  The method for extracting the Stokes $V$
parameter from the data is described in the FORS user manual \citep{FORS} and
by \citet{2002A&A...389..191B}.

Example Stokes V data are presented in Fig.~\ref{fig_FORSV}; in this figure,
and for the field strength determination (as in Fig.~\ref{fig_FORSB}), there
is no rebinning from pixel space to uniform wavelength space. We calculated the
wavelength for the respective data points based on the HgCd and He arc line
spectra taken during daytime. 

Flat fielding is largely unnecessary for the Stokes $V$ determinations,
because the alternate measurement of $-V/I$ and $V/I$ at $-45\degr$ and
$45\degr$ retarder plate angles cancels out any pixel-to-pixel variations in
the CCD response; nevertheless, because intensity ($I$) data are required to
derive the field strength, we applied an appropriate flat-field correction to
all spectra.

\begin{table*}
\centering
\begin{tabular}{lrrrrr} 
 MJD       & \multicolumn{1}{c}{$\langle B_z \rangle$ }  &  \multicolumn{1}{c}{$\sigma_{B_z}$} &
 $W_{\lambda, {\rm He}{\rm I}\,4388}$  & $W_{\lambda, {\rm He}{\rm I}\,4713}$  & phase\\
  & \multicolumn{1}{c}{[G]} &  \multicolumn{1}{c}{[G]} &  \multicolumn{1}{c}{[m\AA]} &  \multicolumn{1}{c}{[m\AA]}\\
\hline
54652.327 &    -2210 &  $\pm$ 130 &  1720  & 530  & 0.681 \\
54656.078 &    -1760 &  $\pm$ 130 &   850  & 310  & 0.875 \\
54656.146 &      230 &  $\pm$ 100 &   740  & 280  & 0.006 \\
54661.327 &    -1110 &  $\pm$ 120 &   660  & 250  & 0.942 \\
54669.186 &      220 &  $\pm$ 110 &   820  & 310  & 0.013 \\
54669.330 &     3200 &  $\pm$ 130 &  1530  & 480  & 0.289 \\
\hline
\end{tabular}
\caption[]{List of magnetic field and equivalent width measurements derived
  from the FORS1 data. The typical error for the equivalent width is about
  5\,\%. For computing the phase we use the ephemeris defined in
  Eqn.~\ref{eqn_eph}. Only Balmer lines were included in the field
  measurements.}
\label{tab_result}
\end{table*}

\section{Results}

\subsection{Magnetic field}\label{sec_mcbs}
As discussed by \citet{2002A&A...389..191B}, under the weak field
approximation one can derive the mean longitudinal magnetic field component
over the stellar disk $\langle B_z \rangle$ from the circular polarimetry and
the gradient of the intensity spectrum with the equation:

\begin{equation}\label{eqn_reg}
\frac{V}{I} = \left\lbrack -{\rm g}_{\rm eff} {\rm C}_z \lambda^2 \frac{{\rm
    d} I}{I~ \rm{d} \lambda} \right\rbrack \times \langle B_z \rangle
\end{equation}

Here we used a factor ${\rm C}_z$ summarizing the physical constants like the
electron charge ${\rm e}$, the electron mass ${\rm m_e}$ and the speed of 
the light ${\rm c}$ in the form:
\begin{equation}\label{eqn_reg2}
  {\mathrm C}_z = \frac{\mathrm e}{4 \pi {\mathrm m}_{\mathrm e} {\mathrm
      c}^2} \sim 4.67\,10^{-13} \mbox{\rm \AA}^{-1 }\mbox{\rm G}^{-1}
\end{equation}
and an effective Land\'e factor ${\rm g}_{\rm eff} = 1$ as discussed by
\citet{1994A&A...291..668C}.

It should be clearly mentioned that there are a number of approximations and
assumptions in equation~\ref{eqn_reg} above. One of those is the week field
approximation for the magnetism, i.e.\ a Zeeman split of less then the
intrinsic line thermal and pressure broadening, which is, however, safely
applicable to a field of only a few kilogauss in a main sequence B star. A
concern prompting far more caution when interpreting the derived numbers is
that the spectra are, in fact, an average over the stellar disk longitudinal
field component, without considering e.g.\ the limb darkening \citep[See][for
  a full discussion of the limitations and caveats of the
  method]{2002A&A...389..191B}. {However, as this quantity, labelled $\langle
  B_z \rangle$, is typially the most easily derived magnetic observable, it is
  also the most published in the literature \citep{2003ASPC..305...16W}, and
  as such easily comparable to other measurements. A more complete magnetic
  modelling to derive the physical dipole field strength is left to a later
  work.}

We used a $\chi^2$ minimization approach to fit the linear model
$V/I=A+\langle B_z \rangle \times x$ to the observational data; this model is
based on eqn.~\ref{eqn_reg} with the addition of a constant term $A$.  {The
  additional term $A$ is physically expected to be zero, as it is the $V/I$ of
  the continuum. However, the employed measurement principle does not
  necessarily guarantee $A=0$ in the measured data, so we allow non-zero
  values in order to improve the linear regression to the slope. It actually
  turns out that the derived values for $A$ are consistently above zero by
  between $0.4\times10^{-4}$ and $2.5\times10^{-4}$ with a typical
  $\sigma=0.5\times10^{-4}$, and on average $1.5\times10^{-4}$. This likely is
  an instrumental effect. In any case this is rather small compared to the
  $V/I$ peak-to-peak amplitude induced by the magnetic field, which is up to
  40 times higher (see Fig.~\ref{fig_FORSV}).}.

The formal errors of $V/I$, which are
derived from the well known photon- and detector noise characteristics, were
used to assign weights for the $\chi^2$ fit. The derived $\langle B_z \rangle$
are given in Table~\ref{tab_result}.

Given recent concerns expressed over the error bars associated with FORS1
magnetic field measurements \citep[see, e.g.][]{2009MNRAS.398.1505S}, we have
exercised particular care in deriving the errors listed in
Table~\ref{tab_result}. Applying the bootstrap Monte Carlo approach described
by \citet{1992nrfa.book.....P}, we used the observational data from each epoch
to construct one million corresponding synthetic datasets. We then applied the
same fitting approach as above to obtain $A$ and $\langle B_z \rangle$ values
for every synthetic dataset. These values exhibit a distribution about those
derived from the actual observations; Fig.~\ref{fig_FORSB} shows this
distribution for the MJD\,54\,669.330 observations, both as a 2-dimensional
probability density map, and as a 1-dimensional probability density function
in $\langle B_z \rangle$. Plotted over the map are contours of constant
$\chi^2$ that enclose 68.3\%, 95.4\% and 99.7\% of the synthetic datasets;
likewise, plotted over the probability density function are the symmetric
bounds enclosing 68.3\%, 95.4\% and 99.7\% of the synthetic datasets. The
bounds for the 68.3\% case are assigned as the error bars quoted in
Table~\ref{tab_result}; because the probability density function in
Fig.~\ref{fig_FORSB} is close to Gaussian (and likewise for the data from
other epochs), these error bars can be regarded as approximate 1-$\sigma$
confidence intervals. Note, however, that we have not made any specific
assumptions regarding the propagation of errors through our modeling process;
the Monte Carlo approach naturally results in error estimates that directly
reflect the combined characteristics of the observations and the modeling.

\subsection{Spectral lines}
Next to the Stokes $V$ spectra, we examined the intensity spectra observed by
FORS1. Due to the very high $S/N$ and the large rotational velocity, even the
relatively low dispersion of FORS1 allows high quality equivalent width (EW)
measurements of the spectral lines.

\begin{figure}
\centering
\includegraphics[viewport=30 190 558 740,angle=0.0,width=\columnwidth,clip=]{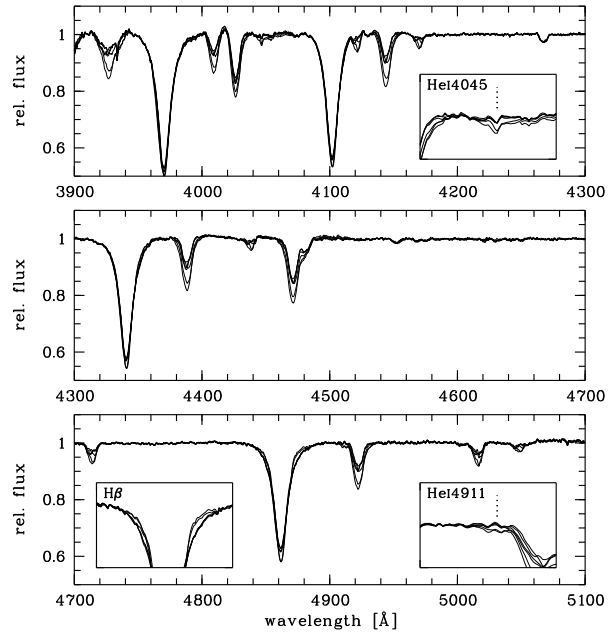}%

\caption[]{Intensity spectra measured by FORS. The {He}{\sc i}-variability is
  well seen, while the Balmer lines vary only little and the {C}{\sc ii}\,4267
  and {Si}{\sc iii}\,4552 lines remain stable.}
\label{fig_FORSI}
\end{figure}

The {H}{\sc i} lines, with the exception of H$\beta$, show little variability,
except in the very core. The spectra taken at MJD=54656.146 and 54669.186 show
enhanced core absorption. At these epochs the $\langle B_z \rangle$ had null-detections, which
is in agreement with the crossing of the magnetic equator through the line of
sight, because this is the time at which a corotating cloud of circumstellar
matter, located at the crossing of the magnetic and rotational equators, is
expected to pass in front of the star.

In H$\beta$ this circumstellar absorption is seen as well, but in addition the
other four spectra show variability in the line wings, in particular in the
red one. This is probably a signature of the circumstellar emission arising in
the corotating clouds, well seen in the two FEROS spectra
\citep{2008A&A...482..255R} for H$\alpha$.

All {He}{\sc i} lines in the observed range follow a similar variation
pattern, varying both in strength and in profile.  The lines showing strong
broadening wings have a larger amplitude in $W_\lambda$ due to variation in
these wings, but the profile variability is better seen in weaker lines, like
{He}{\sc i}\,4713.

\subsection{Forbidden He{\sc i} lines}

In addition to the well known stellar {H}{\sc i} and {He}{\sc i} lines, there
is significant variability at $\lambda \approx 4045$\,\AA\ and bluewards of
{He}{\sc i}4922, at $\lambda \approx 4911$\,\AA. We identify the features with
the forbidden lines {He}{\sc i}4045 ($2^3P\rightarrow5^3P$) and
{He}{\sc i}4911 ($2^1P\rightarrow4^1P$), features
well known in extreme Helium stars \citep{1998ApJ...496..395B}.

{In the FORS data, variability due to forbidden components of He{\sc i} at
  $\lambda\lambda$ 4045 and 4911 was found. At a closer inspection, this
  spectral signature is also present in the old FEROS spectra, and a search in
  other magnetic He-strong stars reveals the presence of these lines also in
  $\sigma$~Ori~E and V\,1046~Ori. Less certain, though not excluded from our
  archival data, are those lines in HD\,64\,740 and HD\,37\,776
  (Fig.~\ref{fig_phases}). These lines are never seen in B stars with normal
  He abundance.

In other words, the presence of these lines indicates a Helium overabundance
wrt. solar values.  Such He-strong stars, especially when having rotationally
broadened lines, are quite hard to diagnose without detailed abundance
analysis, as has also happened for HR7355, instead of classifying it as
chemically peculiar it was rather classified as a later type that it actually
has. This means the presence of the two forbidden Helium lines can safely be
taken as an indicator for a He-strong star. As many He-strong stars are
magnetic (and He-weak stars are typically of later spectral type), with the
same reasoning we consider the presence of these lines as an indicator for a
magnetic field in early B-type stars.}

\begin{figure}
\centering
\includegraphics[angle=270,width=\columnwidth]{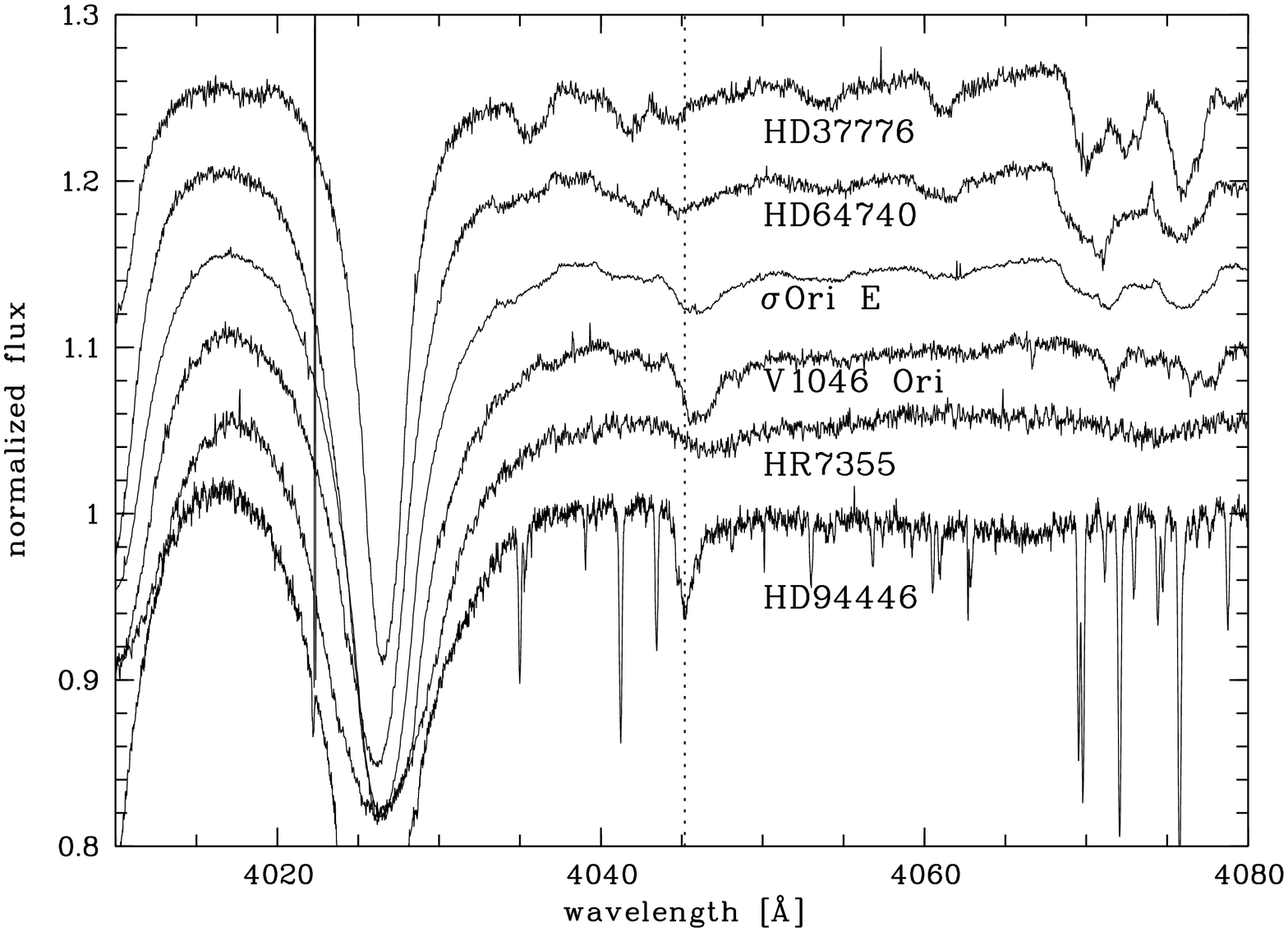}%

\includegraphics[angle=270,width=\columnwidth]{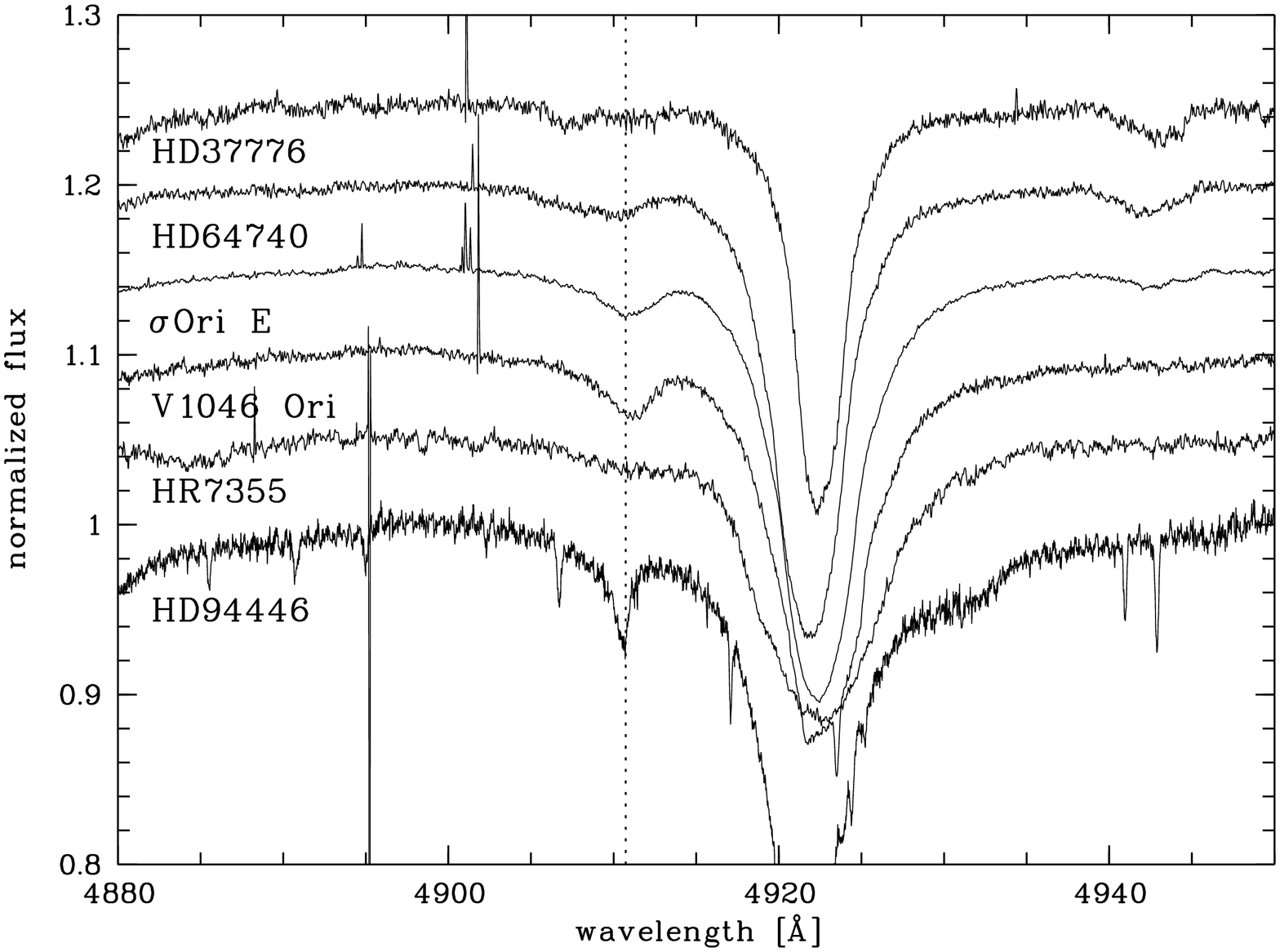}%

\caption[]{Forbidden He{\sc i} 4045,4911 signatures in some known magnetic
  He-strong stars.  The forbidden components are clearly seen in in the slowly
  rotating HD\,94\,446, but also in HR\,7355, V1046 Ori and $\sigma$\,Ori~E,
  and possibly in HD\,64\,740 and HD\,37\,776 as well. Note that these are
  single spectra, so shifts occur as the features are Doppler shifted due to
  the rotational variation.}
\label{fig_forb}
\end{figure}

\section{Discussion}
\subsection{Ephemeris}
{As we know from $\sigma$\,Ori~E \citep{2000A&A...363..585R}, the phase most
  easy to pin down, both spectroscopically and photometrically, is the center
  of one of the two circumstellar clouds crossing in front of the star. We
  chose this point also for HR\,7355 as $\phi=0$. A corresponding epoch $T_0$
  can then be identified as the time of strongest absorption in H$\alpha$,
  i.e.\ the maximal equivalent width. While our FORS data do not cover
  H$\alpha$, and the FEROS data were not taken in this phase, a spectroscopic
  campaign with the echelle instrument UVES at Paranal that has been completed
  in the second half of 2009, suggests MJD$=54\,940.33$ as epoch. However,
  since a full analysis of the UVES data is far beyond the scope of this
  discovery report, and the epoch is not as critical as the period for our
  purpose, we leave a detailed analysis and discussion of the H$\alpha$
  equivalent width curve to a later work (Rivinius et al, in prep.), and here
  just adopt the epoch. Nevertheless it is reassuring that additional
  absorption in H$\beta$ in the FORS1 data is observed at $\phi=0.006$ and
  $0.013$, when the data are sorted with the period derived below.  These two
  occurrences, 25 cycles apart from each other and more than 500 cycles apart
  from the selected epoch, are in full agreement with this epoch derived from
  UVES data.

  For the period, \citet{2008A&A...482..255R} gave a value of 0.521428(6)\,d
  for the Hipparcos data, under the condition that the variations were
  double-wave sinsoidal. The single wave period, which in any case would have
  been to short to be rotational, is firmly excluded by the new magnetic
  measurements.

  { However, also the 0.52\,d period is not able to satisfactorily phase
    all available data, and so its value had to be improved. In order to do
    so, we demand a period to sort all three data types, i.e.\ the photometric
    data (1990-1993, double wave), the magnetic data (2008, single wave), and
    the equivalent widths (1999-2008, possibly double wave). With respect to
    the originally published value, the closest period soring all three
    data-types is $0.521442(4)$\,d. Although the EW data suffers strong
    seasonal aliasing, already its most nearby alias is excluded by the almost
    completely scrambled photometric curve for this period.  The most nearby
    period value for which EW and photometry could be reconciled is
    incompatible with the magnetic phase-curve. We thus conclude that the true
    rotational period of HR\,7355 must be within the above value's
    uncertainties.}


  In a final step, we can assume that the photometric minima do have a certain
  phase relation to the spectroscopic curve. {There are two possiblities:
    First, the minima could be due to cloud eclipses in front of the star,
    then we can further require that one of the photometric minima occurs at
    phase $\phi=0$. Second, the photometric might be due to photospheric flux
    modulation in the He-enriched parts
    \citep{2007A&A...470.1089K,mikulasek}. In this case, the photometric
    minima would coincide with minimal {He}{\sc i} equivalent width. However,
    in this particular star, the maximal $H\alpha$ equivalent width and the
    minimal {He}{\sc i} equivalent width are almost simultaneous, so that a
    final decision about this can only be made with new photometry more
    simultanous with recent observations.}  in any case, already with the
  period derived from the equivalent widths alone one of the two photometric
  minima is very close to $\phi=0$, so that under the assumption that the
  photometric variability is due to the circumstellar material obscuring the
  line of sight the period becomes $P=0.521444(3)$\,d. We will use the latter
  value for the discussion, but note that this relies on the assumption that
  the photometric minima are due to eclipses, while the best period without
  this assumption is $P=0.521442(4)$\,d

  The ephemeris used thus is
  \begin{equation}\label{eqn_eph}
T_{\rm {H}{\rm I}~max.~absorption}({\rm MJD}) = 54940.33 + 0.521444(3) \times E
\end{equation}

\subsection{Periodic variations}\label{subsec_per}
  With the above choice of epoch we expect to see photometric minima, as well
  as $B_z=0$, at phase $\phi=0$, i.e.\ when the magnetic equator is facing towards us
  (Fig.~\ref{fig_phases}). Then using the above ephemeris, and assuming a
  sinusoidal variation of the magnetic field, we estimate the field curve as
\begin{equation}\label{eqn_field}
\langle B_z \rangle(t) = 350\,{\rm G} + 2850\,{\rm G} \times 
\sin\left(2\pi \left(\frac{
    t \rm -T_0}{P} - 0.02\right)\right)
\end{equation}
  where $t$ is the date, and $T_0$ and $P$ are epoch and period from
  Eqn.~\ref{eqn_eph}, and the shift of 0.02 in phase is required to fulfill
  the above condition $B_{z,\phi=0}=0\,$G, due to the constant term of 350\,G.

\begin{figure}
\centering
\includegraphics[viewport=0 215 543 745,width=\columnwidth,angle=0.0,clip=]{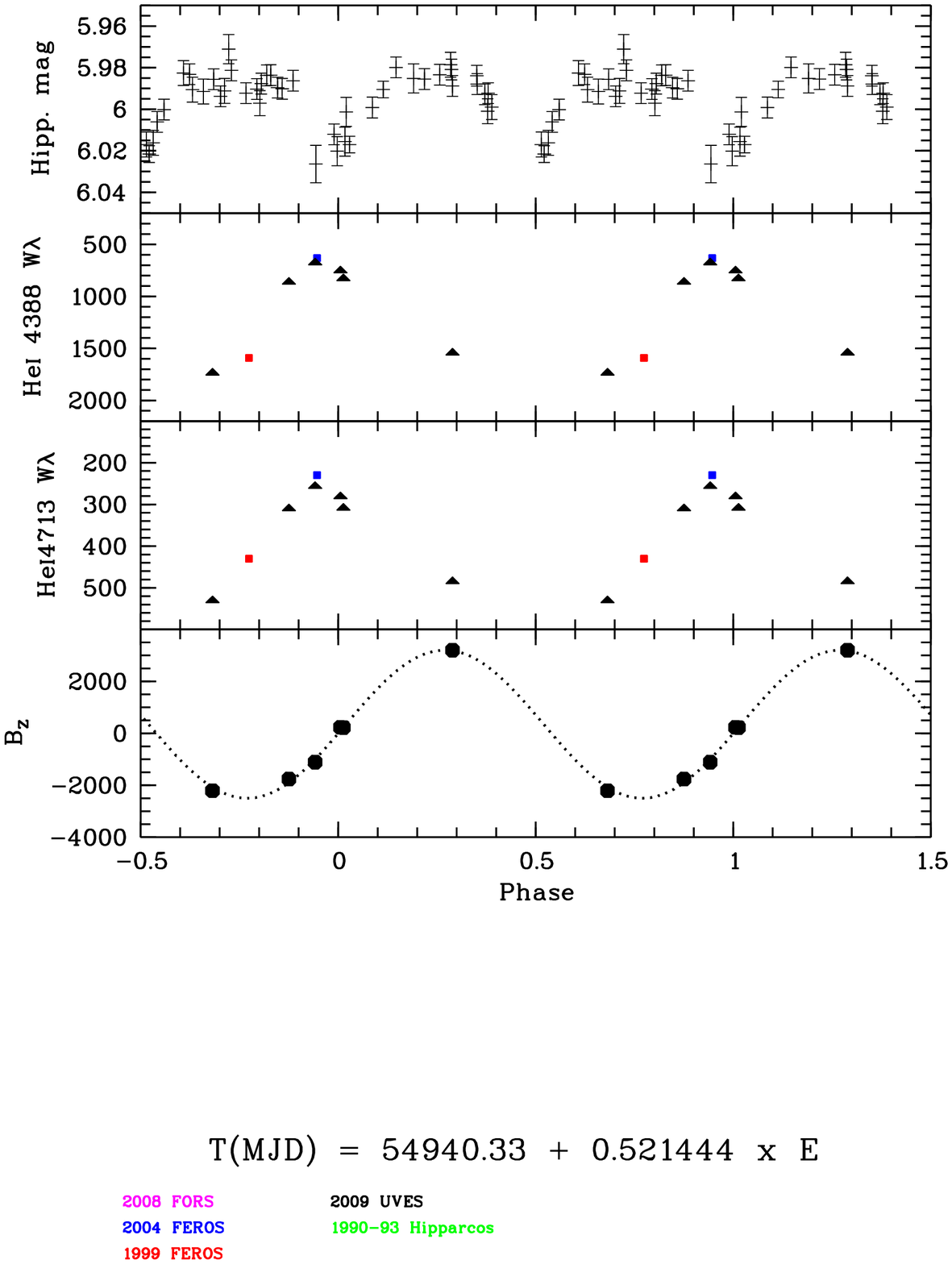}
\caption[]{Measured quantities (see Table~\ref{tab_result}, also for typical
  errors) as function of the rotational phase (Eqn~\ref{eqn_eph}). The
  Hipparcos double wave light-curve is shown in the uppermost panel, the
  {He}{\sc i}\,4388 and 4713 equivalent widths below as triangles. The
  $W_\lambda$ measured in the FEROS spectra \citep{2008A&A...482..255R} are
  shown as red (1999 data) and blue (2004 data) squares. The lowermost panels
  shows the $\langle B_z \rangle$ field measurements and the curve defined by
  Eqn.~\ref{eqn_field}. }
\label{fig_phases}
\end{figure}
The fact that $B_{\rm max}\approx - B_{\rm min}$ confirms the assumption of
$i\approx 90^\circ$ by \citet{2008A&A...482..255R}, while the term of 350\,G
is easily explained either by a slight offset from this value, or by an
off-center magnetic dipole. 

Such an off-center dipole would cause a non-sinusoidal field curve. {In
fact, due to two datapoints being taken at identical phases, there are
effectively only five points to constrain the three free parameters of
Eqn.~\ref{eqn_field}.  The field curve, although it seems to fit very well, is
thus not well constrained, which is why we cannot really give
confidence limits for the parameters unless further data has been obtained. In
particular, we stress that the exactly sinusoidal shape of the curve is rather
an assumption than an actually observed property.}}

When the magnetic poles face the observer, $\langle B_z \rangle$ becomes
maximal. Although there are only few points, not sampling the curve in all
detail and in particualr not necessarily the respective maxima and minima, it
is clear that the {He}{\sc i} absorption is much stronger in these phases than
when the magnetic equator is visible (Fig.~\ref{fig_phases}). This is in full
agreement with the behavior observed in other He-strong stars, like
$\sigma$\,Ori\,E \citep{2000A&A...363..585R}.
 
{The amplitude of the {He}{\sc i} EW variations is considerably larger
  than in $\sigma$\,Ori\,E, however. There the maximal EW is only about a
  factor of 1.3 to 1.5 stronger than the minimal one, depending on the
  spectral line. In HR\,7355 the lines strengthen, wrt.\ their minima, by a
  factor of 2 for lines like {He}{\sc i}4713, and even a factor of 3 for
  strong lines with significant broadening wings like {He}{\sc i}4388. This
  strong modulation is indicative for two large Helium enhanced patches on the
  surface close the equator at opposite longitudes, which point to a large
  angle $\beta$ between the rotational and magnetic axes.}

\section{Summary}

The results confirm the magnetic and He-strong nature of HR\,7355. We have
detected a magnetic field of multi-kilogauss strength, varying with the
rotational period, {which is $P= 0.521444(3)$\,d under the assumption of
  the photometric minima being eclipses, and $P= 0.521442(4)$\,d without this
  assumption.}
In order to avoid a potential underestimation of the confidence limits of the
magnetic field measurement \citep[as suspected for previous FORS measurements
  by][]{2009MNRAS.398.1505S}, we derived them with a boot-strap Monte-Carlo
method, which does not implicitely assume a statistics for the error
propagation, but numerically reconstructs the probability distribution from
which the actual observation was drawn.

The observed magnetic field and the suggested topology (see
Sect.~\ref{subsec_per}) is in agreement with the initial hypothesis by
\citet{2008A&A...482..255R}, namely that of an equatorially seen star,
i.e.\ $i\approx 90^\circ$, with an oblique magnetic dipole. This topology
gives rise to a double-wave light curve, either as the corotating
magnetospheric clouds, magnetically bound at the crossings of magnetic and
rotational equator, sweep through the line of sight twice per rotational
period \citep{2008MNRAS.389..559T}, or due to the modulation of the
photospheric flux by the abundance pattern \citep{mikulasek}.

\section*{acknowledgements}
We thank G.~Wade and D.~Bohlender for discussions, and making us aware of
potential traps to be avoided.

\bibliographystyle{mn2e}
\bibliography{7355a_mn,7355b_mn}

\clearpage

\end{document}